# Quasi one-Dimensional Band Dispersion and Metallization In long Range Ordered Polymeric wires

G.Vasseur[1], Y. Fagot-Revurat[1*], M. Sicot[1], B. Kierren[1], L. Moreau[1], D. Malterre[1]

[1.] *Institut Jean Lamour, UMR 7198, Université de Lorraine/CNRS, BP 239 FE-54506, Vandoeuvre-les-Nancy, France*

L. Cardenas[2], G. Galeotti[2], J. Lipton-Duffin[2], F.Rosei[2]

[2.] *Centre Énergie, Matériaux et Télécommunications, Institut National de la Recherche Scientifique, 1650 Boulevard Lionel-Boulet, Varennes, QC, J3X 1S2, Canada*

M. Di Giovannantonio[3], G. Contini[3].

[3.] *Instituto di Struttura della Materia, CNR, Via Fosso del Cavaliere 100, 00133 Roma, Italy*

P. Le Fèvre[4], F. Bertran[4].

[4.] *Synchrotron SOLEIL, L'Orme des Merisiers, Saint-Aubin, Gif sur Yvette, France*

L. Liang[5], V. Meunier[5]

[5] *Department of Physics, Applied Physics, and Astronomy, Rensselaer Polytechnic Institute, New York 12180, United States*

D. F. Perepichka[6],

[6.] *Department of Chemistry, McGill University, 801 Sherbrooke Str. West, Montreal,QC H3A 0B8, Canada.*

**ABSTRACT**:

We study the electronic structure of an ordered array of poly(para-phenylene) chains produced by surface-catalyzed dehalogenative polymerization of 1,4-dibromobenzene on copper (110). The quantization of unoccupied molecular states is measured as a function of oligomer length by scanning tunneling spectroscopy, with Fermi level crossings observed for chains longer than ten phenyl rings. Angle-resolved photoelectron spectroscopy reveals a graphene-like quasi one-dimensional valence band as well as a direct gap of 1.15 eV, as the conduction band is partially filled through adsorption on the surface. Tight-binding modelling and ab initio density functional theory calculations lead to a full description of the organic band-structure, including the k-dispersion, the gap size and electron charge transfer mechanisms which drive the system into metallic behaviour. Therefore the entire band structure of a carbon-based conducting wire has been fully determined. This may be taken as a fingerprint of π-conjugation of surface organic frameworks.

A major challenge in modern surface science is to create ordered arrays of covalently linked organic nanostructures. By doping molecular electronic bands into highly conductive states, these structures may be promising for use as elementary building blocks in electronic carbon-based molecular devices such as organic field-effect transistors[1], light emitting diodes,[2,3] photovoltaics,[4] and sensors.[5] Despite its exceptional physical properties, the graphene's lack of a band gap severely limits its potential for creating such devices. Engineering the gap in graphene by using nanostructuring, e.g. creating graphene nanoribbons (GNRs) of narrow width, has been proposed as a feasible route toward carbon-based electronics. Thus, the GNRs' bandgap can be tuned by altering their lateral size or by modifying their edge termination (armchair versus zigzag)[6-8]. An emerging bottom-up approach for producing such carbon nanostructures, exploits covalent linking (polymerization) of precursor molecules on metal surfaces.[9-20] In these materials, functional properties, including the geometry and the band gap, can be tailored by means of a judicious choice of monomer and supporting surfaces.[21-23] The on-surface polymerization is typically demonstrated by measuring the periodicity of polymeric architectures using scanning tunnelling microscopy (STM).[24] Evidence of π-conjugation was shown by combining X-ray photoelectron spectroscopy (XPS) and near-edge x-ray absorption fine structure (NEXAFS).[25] Band gaps can be deduced by scanning tunnelling spectroscopy (STS) and/or angle-resolved photoelectron spectroscopy (ARPES) and supported by theoretical calculations.[26-30] However, a full band dispersion in polymeric chains has not been reported to date, due to the difficulty in obtaining ordered phases at sufficiently long range.

In this work, we unambiguously establish the full band structure of a surface-confined π-conjugated organic polymer, as well as the impact of the substrate on its electronic properties. A long-range ordered array of poly(para-phenylene) (PPP) chains was produced through the surface-catalysed dehalogenative polymerization of 1,4-dibromobenzene (dBB) on copper (110). The high structural quality of the molecular layer, combined with the large extent of the individual PPP oligomers permitted both local and surface-averaged studies. Energy-dependent standing wave patterns observed by STS in finite-size PPP oligomers allowed the determination of the k-resolved conduction band dispersion. This one crosses the Fermi level, conferring to the polymer a metallic character. Using ARPES, we measured the full valence band which disperses along the chains on 6.7 eV, crossing the bulk states of the substrate. As the conduction band is partially occupied, a 1.15 eV band gap was directly observed. A Hückel tight-binding (TB) model provides understanding of both ARPES and STS measurements, allowing the estimation of both effective intra- and inter- chains resonance integrals and establishes the quasi one-dimensional nature of the dispersion. DFT calculations fully reproduce the band structure and point out a strong hybridization at the organic/metal interface which is responsible for filling of the polymer unoccupied states.

**A long range ordered and commensurate PPP polymeric phase on Cu(110)**

A systematic investigation of the dBB/Cu(110) interface as a function of coverage and annealing temperature allows us to identify an unreported structural arrangement, which was used as a starting point for the formation of an ordered polymer phase via thermal activation. In a previous work, Di Giovannantonio *et al* demonstrated that the sublimation and thermal treatment of dBB on Cu(110) leads to the formation of PPP chains.[25] This process is understood to be an Ullmann coupling reaction, which is summarized in Figure 1a: depositing the molecules at room temperature leads to an organometallic (OM) phase[25,31], where Br-C bonds are replaced by C-Cu-C bridges. Subsequent annealing above 460 K promotes C-C bonding, resulting in the formation of coaligned π-conjugated PPP chains with a lattice parameter of 4.4 Å as measured by STM. In that case, the PPP chains were aligned along the <1-10> direction, for which the interatomic Cu distance of the surface is 2.55 Å. This leads to the incommensurability between the polymer and the substrate (see section 2 and figure S3, SI). Probably because of this lattice mismatch, the length of defect-free oligomer units tends to be limited. However, a close inspection of the growth reveals the apparition of an additional OM phase at high coverage, above 0.9 monolayer (ML), which co-exists with the OM phase described above and fully dominates the surface at 1 ML. STM images and low energy electron diffraction (LEED) patterns characterising these two phases are provided and discussed in the Supplementary Information (see section 1 and figure S1 and S2, SI). When this high coverage phase is annealed at 500 K, alternating rows of polymers and bromine atoms are formed, as shown in the STM images presented in Figure 1b. The measured internal periodicity of 4.4±0.2 Å agrees with the expected phenyl-phenyl spacing characterizing the PPP. The polymer chains produced from high coverage are packed into domains and aligned along the two <1-1-2> and <1-12> substrate directions, a detail which complicates measurement by surface-averaging techniques such as photoelectron spectroscopy. While a single domain orientation dominates in figure 1b, both orientations are observed in longer-range STM images, and cover the surface with equal probability (see figure S4, SI). The substrate periodicity along <1-12> is 4.43Å, which matches with the lattice parameter of the PPP. Thus, the polymer is here fully commensurate with the surface and may be grown to long size with little or no strain arising from lattice mismatch (see section 3, SI).

The results of DFT optimization of the surface structure are presented in Figure 1c (see method section for details).[32-34] The PPP polymer (C atoms are in grey) is adsorbed in a flat geometry, 2.2 Å above the copper surface. Such a short distance (cf. 3.35 Å for inter-planar spacing in graphite) points to a significant hybridization of the electronic levels of the polymer with the copper interface states. Bromine atoms (green spheres) are adsorbed on the short bridge (SB) site of the underlying Cu lattice (orange spheres). The center of each phenyl ring overlaps the hollow (H) site, which is the preferred site for benzene adsorption on Cu(110).[35] A STM image calculated in the Tersoff-Haman approximation[36] at a bias voltage of 200 mV is an excellent match to the experimental data (Figure 1d and 1e).

**Local spectroscopy and tight-binding modelisation.**

Using STS, confinement of unoccupied molecular states has been observed previously in individual polythiophene chains,[37] and more recently, in 7-AGNRs[30] and GNR heterojunctions.[38] We recorded differential conductivity maps as a function of chain length, in order to build the k-resolved band structure associated to the conduction band of the infinite polymer. STS measurements obtained on PPP oligomers with N=6 (26 Å), N=10 (44 Å) and N=22 (97Å) phenyl rings are presented in Figure 2a. As expected, the local density of states (LDOS) is spatially modulated along the chains, leading to standing wave patterns which depend strongly both on the bias voltage and the oligomer length. For N=6, the lowest unoccupied molecular orbital (LUMO) (indicated by a blue segment in figure 2a) is slightly above the Fermi level. As the length of the chain is increased, the LUMO progressively moves below the Fermi level, reaching -0.4 eV for the N=22 polymer. The energy separation between the different states is also reduced until it becomes impossible to distinguish them for long polymers (>N=20).

A simple TB model (see method section for details) reproduces our results. The long range modulations observed in the LUMO, LUMO+1 and LUMO+2 orbitals are in agreement with the local conductance maps recorded for the first three states in the polymer chains, as shown in Figure 2b,c for the N=6 case. Using a single C-C resonance integral β=-3.6 eV, an excellent fit to experimental binding energies as a function of oligomer length is observed as presented in Figure 2d. Defining $k_p=2p\pi/Na$ (p<N, a=4.4 Å) as the wave vector associated to the 1D-confined states allows us to build the conduction band dispersion for the infinite polymer, as shown in Figure 2e. A parabolic fit (black dashed line) to the bottom of the conduction band implies an effective mass of 0.16 $m_0$ for charge carriers in the vicinity of the Fermi level. This value is in agreement with those obtained for AGNRs.[30] As the band crosses the Fermi level at $k_F=\pm0.12$ Å$^{-1}$, the PPP polymers are metallic and thus represent an array of conducting nanowires. None of the occupied states were resolved by STS, and thus we were unable to extract the band-gap of the PPP.

**Band-structure from angle-resolved photoelectron spectroscopy.**

ARPES has previously been used to reveal quantum well states in molecules[39,40] or to reconstruct the discrete electronic orbitals of self-assembled individual molecules at surfaces.[41,42] Direct evidence of a band-structure associated with the C-C covalent bonding and a long range delocalization of charge carriers have been reported to date only on graphene materials.[26] Here, ARPES intensity maps were measured along the <1-12> axis, parallel to the chain direction of one domain (therefore probing roughly half of the monolayer). The incident photons were p-polarized, with an energy of hν=35 eV. The spectral weight from the second domain, which is rotated by 70.52° with respect to the first, does not contribute to the ARPES signal because its k-points are not accessible in this experimental geometry, except at the Γ point. Comparison of surfaces both with and without polymers permits the identification of a strongly dispersive band, that crosses the 3d states of the substrate, labelled in Figure 3a using yellow arrows. The bottom of the band is located at the Γ point at $E-E_F=-8.1$ eV, whereas the top of the band reaches a maximum at $E-E_F=-1.4$ eV at $k_{//}=1.4$ Å$^{-1}$. The absence of this band in the clean Cu(110) spectra confirms its molecular origin (see Figure S5, SI). In addition, a strong decrease of the ARPES signal is observed when the photon polarization vector is changed from p-polarization (having an out of surface plane component of the polarization vector parallel to π orbital axis) to s-polarization (having only in surface plane component of the polarization vector) consistent with the π-character of the molecular orbitals and the flat-lying geometry of the phenyl rings (see Figure S5, SI). This molecular band is therefore identified as built from the highest occupied molecular orbitals (HOMOs).

Since the size of the first Brillouin zone (BZ) is 1.43 Å$^{-1}$ (2π/d, using the distance measured by STM d=4.4 Å), the apparent periodicity of the band (2.88 Å$^{-1}$) does not seem to match the PPP periodicity. However, calculation of both band structure and ARPES spectral intensity using the Hückel (Tight-Binding) models provides a full understanding of this phenomenon. The valence band structure of the infinite polymer, calculated in the TB model using a resonance integral of β=-3.5 eV, is presented in figure 3b (red dashed lines). Here the molecular band was rigidly shifted by 0.1 eV to match the data. Three molecular bands are easily identified, two of them dispersing in opposite phase and crossing at the edge of the first BZ i.e. for $k_{//}$=±π/d=0.71 Å$^{-1}$, in respect of the periodicity of the system. However, the theoretical ARPES intensity (colorscale on the figure 3b) was deduced from the Fourier-transform of each molecular orbitals calculated for a N=20 polymer, according to the Fermi's golden rule as proposed in the cases of sexyphenyl on the Cu(110) surface[41] and PTCDA on the Ag(110) surface[42] (see section 5 and figures S6 and S7, SI). It appears that, as usually oserved in ARPES, the spectral intensity does not possess the periodicity of the system, due to cross-sectional effects in the photoelectron emission. This result is in agreement with the experimental measurement presented in figure 3a. Therefore, knowledge of the orbital topology and a rudimentary model of the electronic structure are necessary to explain the appearance of experimental dispersion curves.

High-resolution measurements taken in the 2$^{nd}$ BZ close to the top of the occupied molecular band show that the molecular spectral weight disperses up to a binding energy of -1.4 eV at $k_{//}$=1.42Å$^{-1}$, while a small portion of a higher energy band dips below the Fermi level. A direct band gap of 1.15 eV is found between the minimum of this band and the top of the valence band described above. This result is in agreement with the location of the conduction band as revealed by our STS measurements on long PPP polymers, confirming the metallic character of the polymer.

In addition, a constant energy map taken at E-$E_F$=-1.8 eV, close to the molecular band maximum, is shown in Figure 3d (see also Figure S8, SI). Two almost completely linear contributions are clearly identified at well-defined $k_{//}$ values. However, detailed investigation shows that the perpendicular dispersion is not precisely zero, as a periodicity of $k_\perp$=0.6 Å$^{-1}$=2π/b (where b=10.4 Å, the distance between adjacent PPP chains) is apparent in figure 3f. A TB model incorporating an effective inter-chain hopping integral β'=-0.15 eV permits the quantification of this small dispersion (see also section 6, SI). A weak transverse dispersion such as this may correspond to indirect hopping mediated by the substrate and/or by Br atoms.[43] We thus extract an experimental ratio β'/β≈0.04, that allows us to conclude on the quasi one-dimensional nature of the molecular bands.

**DFT calculations and discussion**

In figure 4a (right part), we show that our local STS measurements (blue squares) for different chain lengths are in remarkable agreement with the k-resolved conduction band structure obtained from ARPES (green squares). Red solid lines present the band structure calculated in the TB model using a single C-C resonance integral β=-3.5 eV in agreement with the expected value for graphene-like materials[44]. However, the valence and conduction bands are independently shifted in energy, since our TB model leads to a band gap of 2.9 eV. So far, the precise value of the gap (1.15 eV) and the partial filling of the unoccupied molecular states cannot be understood in the framework of this simple yet powerful method.

DFT computation of the electronic band-structure for the model appearing in Figure 1c are depicted at the left Figure 4a. The bands originating from the occupied and unoccupied molecular states can be distinguished from those originating from the copper substrate by plotting the projection of the band structure onto the top layer of the system, which contains only H, Br and C atoms. The result of this projection is shown as red circles in Figure 4a. The conduction band as well as the bottom and top portions of the valence band can be clearly identified. However the projected band-structure contains several signatures of hybridization with the substrate, which are not captured in the TB model. A comparison between the measured and calculated electronic density of states (DOS) is shown in Figure 4b. The DOS of pristine Cu(110) (filled blue line) is flat between 0 and -2 eV, and originates from copper s-p bands. The additional spectral weight arising from the PPP overlayer appears clearly as peaks corresponding to the top and bottom of the valence and conduction molecular bands, respectively. The separation between these peaks corresponds to a band-gap of 1.15 eV. The calculated gap is approximately 0.90 eV, which is slightly smaller than the experimental value, but this is a common drawback of DFT calculations, which neglect correlation effects. Overall, both the size of the gap and the minimum of the valence band below the Fermi level are captured by these simulations, in agreement with the finding of metallic character in the PPP nanowires.

The gap for freestanding PPP is expected to be 2.9 eV from TB approach[45]. Assuming the PPP polymer as being a 3p-type AGNR with p=1 and w=2.4 Å (see section 7 and figure S$_9$, SI), we expect a band gap of 2.45 eV from DFT calculations corrected at 4.1 eV including GW corrections.[6] We have also carried out DFT calculations in the gas phase using B3LYP/6-31g(d) that yield a band gap of 3.05 eV (see section 8, SI). In another study, STS measurements were performed on isolated PPP chains grown on Cu(111) using a different precursor.[27] The bottom of the conduction band was found to be 0.9 eV above the Fermi level for long chains, in the unoccupied electronic states, i.e. shifted by 1.3 eV compared to our measurements. Thus, understanding the interaction between the polymer and the substrate appears to be of major importance in order to explain both the band gap closing and the filling of unoccupied molecular states as emphasized very recently.[29]

Figure 5a dissects the essential mechanisms that should be taken into account in the energy level alignment and gap modifications in organic semiconductor/metal interfaces. Considering the PPP polymer is adsorbed onto a surface, the resulting position of the molecular levels depends on the substrate work function as indicated on the left side of Figure 5a. The Cu(111) workfunction (4.94 eV) is larger than that of Cu(110) (4.48 eV). Consequently molecular levels are shifted to lower binding energies on the Cu(110). This effect may be enhanced by an additional reduction of the work function $\Delta\phi_s$ due to the compression of the electron cloud inside the metal substrate induced by the proximity of the molecular layer.[46-48] In our case, the short distance of 2.2 Å between the polymer and the surface may explain the large shift observed in the present work as compared to the report on Cu(111). However, it has also been shown that certain physisorbed molecules experience Fermi level pinning: in this case, charge transfer from the substrate to the LUMO creates a new surface dipole which increases the work function until the LUMO lies with the Fermi energy, inhibiting its occupation.[47-49] For chemisorbed molecular materials, substantial molecule/substrate hybridization leads to a direct charge transfer which favours LUMO occupation. This mechanism has been recently identified as responsible for the metallization of pentacenequinone and pentacenetetrone molecules adsorbed on coinage metal surfaces.[29] Furthermore, several works have also reported a strong reduction of the HOMO-LUMO gap induced by the proximity of molecules with a metallic substrate, due to charge screening effects.[48,49] In order to test the impact of the substrate on electronic properties, the band structure was recalculated in a geometry where the separation between the PPP and the substrate is artificially increased from 2.2 Å to 8 Å (Figure 5b). It is clear that: (i) the valence and conduction molecular bands are better defined, having lost the signatures of hybridization with the substrate; (ii) the full molecular band-structure is shifted upwards toward unoccupied states, with the top of the valence band grazing the Fermi level; (iii) the gap has increased to 1.6 eV but still remains much lower than expected from gas phase calculations.[6] In general local density approximation (LDA) and generalized gradient approximation (GGA) underestimate band gaps of GNRs and organic polymers in gas phase as compared to GW calculations[6]. Nevertheless, the new position of the conduction band is close to 1 eV as observed for PPP/Cu(111)[27]. We therefore conclude that hybridization between molecular π-states and copper states is the essential mechanism that permits partial filling of the LUMO, the lowering of the band gap and results in metallic behaviour for the PPP nanowires. This effect is enhanced by the fact that PPP polymers are aligned in such a way as to be commensurate with the Cu(110) substrate, which maximizes here the molecule/surface interaction in comparison to the PPP/Cu(111) interface.

**Conclusion**

We have combined molecular self-assembly and covalent bonding processes to prepare long range ordered arrays of PPP and characterised their full π-band structure. We have demonstrated the quasi-1D metallic character of these chains grown in a commensurable way on the Cu(110) surface. A simple Hückel tight-binding provides a good understanding of both ARPES and STS measurements and allows estimating effective inter- and intra-chains resonance integrals. Beyond, DFT calculations have evidenced the strong hybridisation existing between the polymer and the substrate that explains both band gap reduction and carrier doping. Despite the substantial hybridization, bands characterizing the PPP wires remain strongly quasi-one dimensional leading to conducting polymeric (nano)wires. A future systematic investigation of electronic properties as function of the chemical nature of the substrate and its geometry, as well as role of the halogen species could be relevant in order to tune the band gap and the metallicity. This work is of particular relevance in the framework of covalent organic nanostructures design for potential applications in nanoelectronic devices. Thus, the complete band dispersion of the polymer may be considered as the spectroscopic fingerprint of the polymerisation process of such molecules as DBB on surfaces where the Ullman reaction can be achieved like gold, silver or copper and also as a recent promising way on semi-conductors.[50]

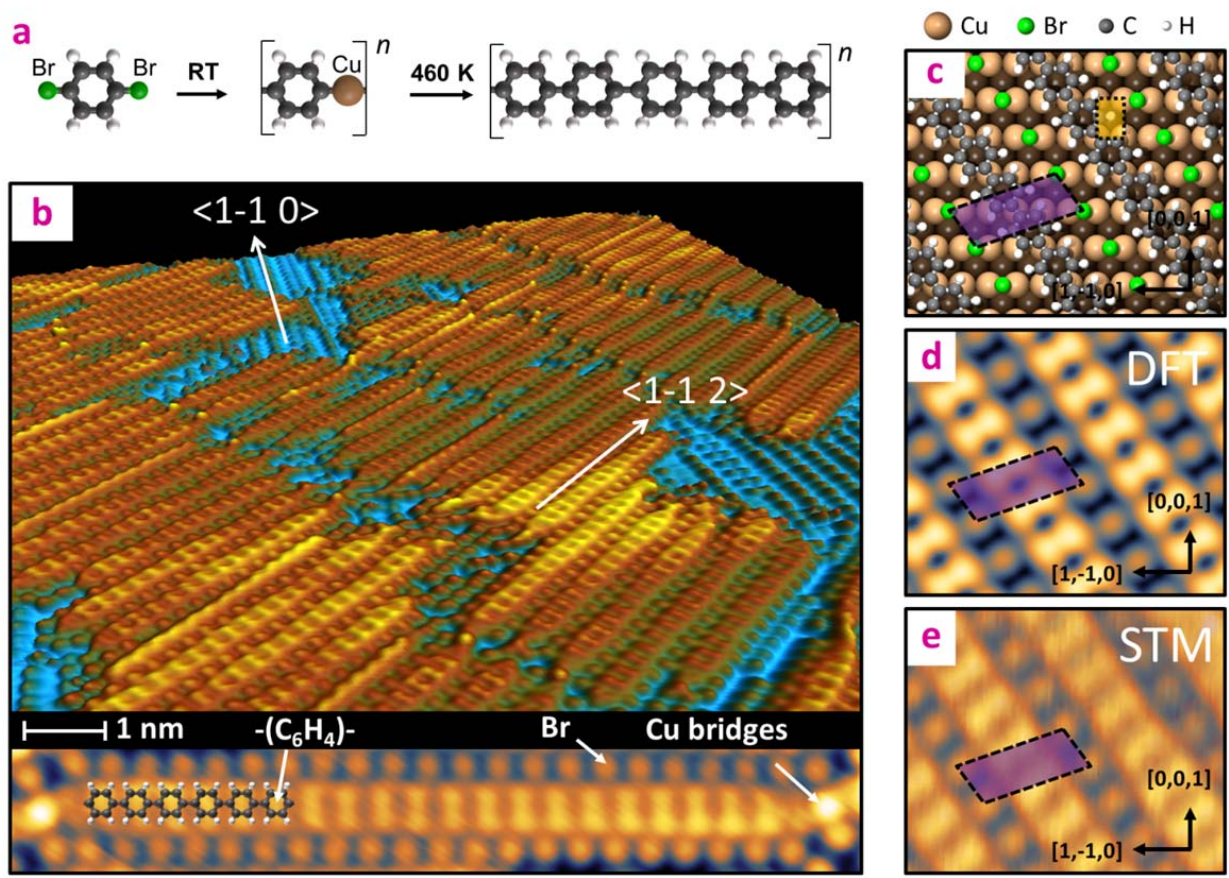

**Figure 1:** (a) Ullmann coupling reaction for 1,4-dibromobenzene on Cu(110); two additional Br atoms linked to the metallic surface (not shown here for clarity), are produced by the Ullmann reaction as experimentally observed. (b) 3D rendering STM image of the poly(para-phenylene) chains grown on Cu(110) for coverage close to 1 monolayer after annealing to 500K (I=0.2nA, U=0.1 V). Blue area contains additional Br atoms adsorbed on the Cu substrate. The bottom part shows the detailed structure of the PPP chains separated by bromine lines; (c) DFT-optimized geometry of the polymerized system. Atoms are represented as spheres according to the legend. The orange and violine areas are the Cu(110) and PPP/Cu(110) surface unit cells respectively. Simulated (d) and experimental (e) STM images of the surface recorded at +0.2eV.

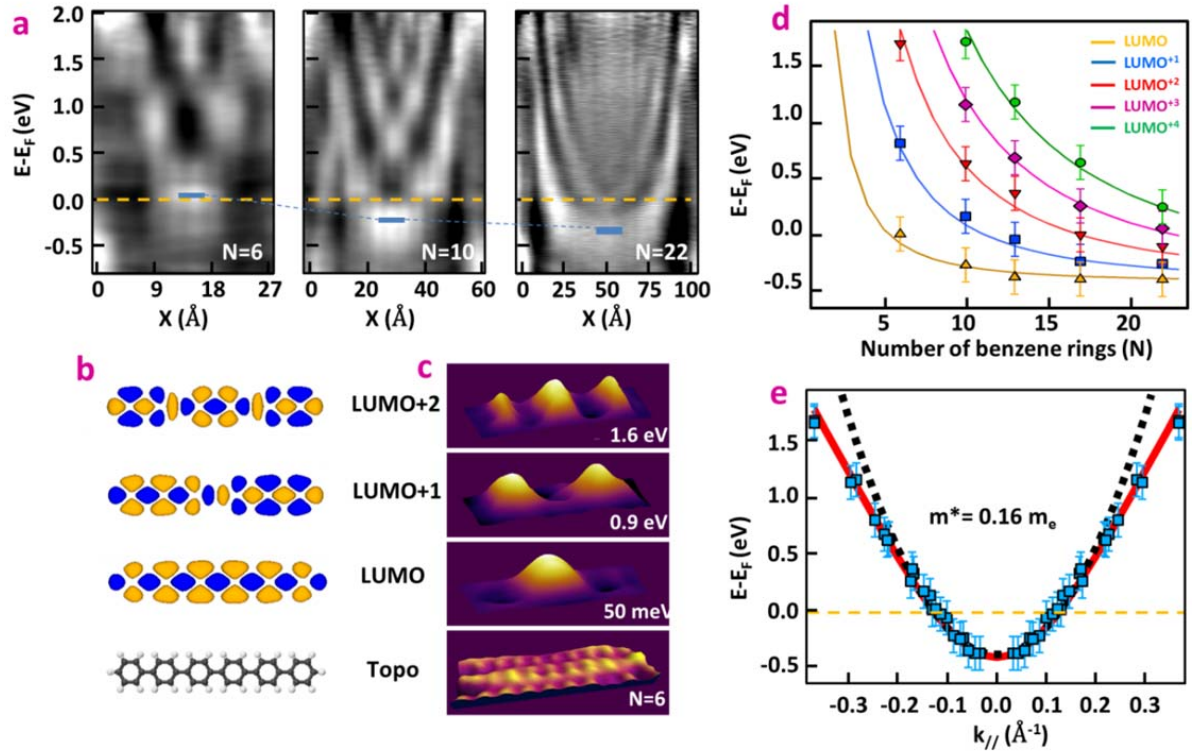

**Figure 2:** (a) experimental dI/dV measurements recorded along a polymer for three lengths (N correspond to the number of phenyl rings). The vertical axis is energy, the horizontal axis is position along the polymer. The blue marks show the bottom of the bands. (b) Schematic representation of the free sexiphenyl molecule and its LUMO, LUMO+1 and LUMO+2 states calculated using a tight-binding model built from Hückel orbitals. (c) Experimental topography (bottom) and STS conductance maps for the N=6 case. (d) Experimental (dots) and theoretical (solid lines, extracted from Hückel's model) evolution of the energy of the LUMO states as function of the length of the polymeric chains. (e) Experimental and theoretical dispersion of the LUMO conduction band. Experimental points (blue) are extracted from data shown in (d), using the relation $k_{//}=2\pi p/Na$ (p=1,2,…). The theoretical band (red solid line), is obtained from a tight binding model using the same resonance integral as the Hückel model. The black dashed curve represent a parabolic dispersion with $m^*=0.16m_0$.

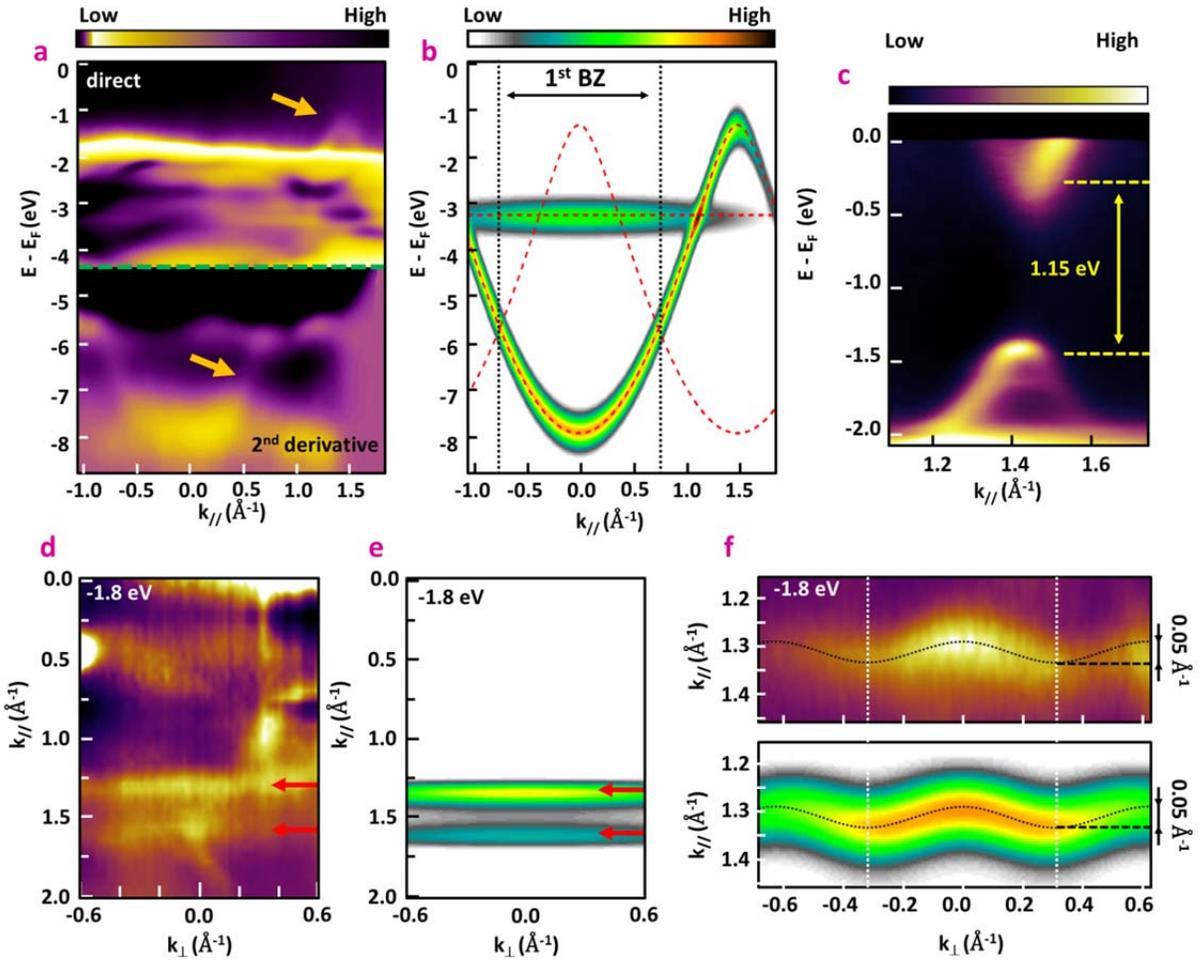

**Figure 3:** (a) ARPES intensity map measured in the <1-12> direction, parallel to the polymers chains for 1ML-PPP/Cu(110). The bottom part is displayed in 2$^{nd}$ derivative. (b) Tight-binding modelling of the band-structure of the N=20 PPP polymer using β=-3.5 eV. (c) High-resolution ARPES intensity map recorded close to the Fermi level showing the HOMO-LUMO band gap. Measurements (d) and tight-binding modelling (e) of the ARPES constant energy map at -1.8 eV. (f) zoom on the measured (top) and calculated (bottom) ARPES constant energy maps. The transverse resonance integral used in the calculations is β'=-0.15 eV.

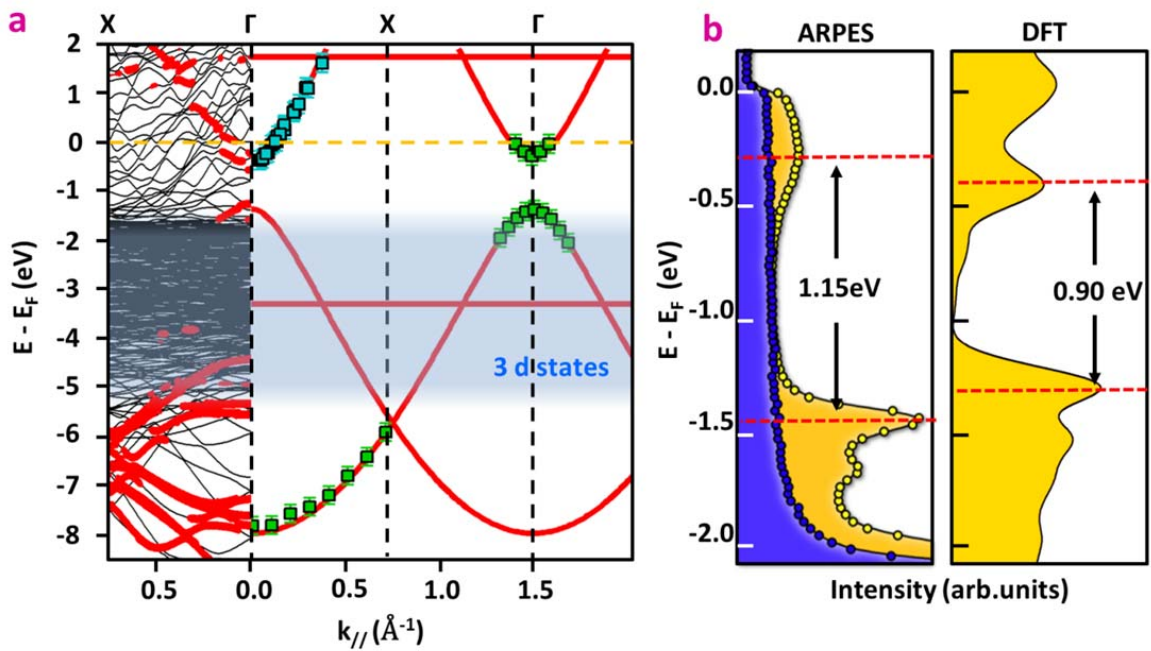

**Figure 4:** Band-structure of PPP polymers grown on Cu(110). (a) DFT band-structure (black lines, red circles correspond to the projection on the molecular layer), tight-binding modelling (red solid lines), ARPES (green squares) and STS (blue squares) experimental dispersion curves. (b) k-integrated PES DOS on the Cu(110) substrate (left panel, blue), on PPP/Cu(110) (left panel, yellow) and corresponding DFT DOS (right panel, yellow).

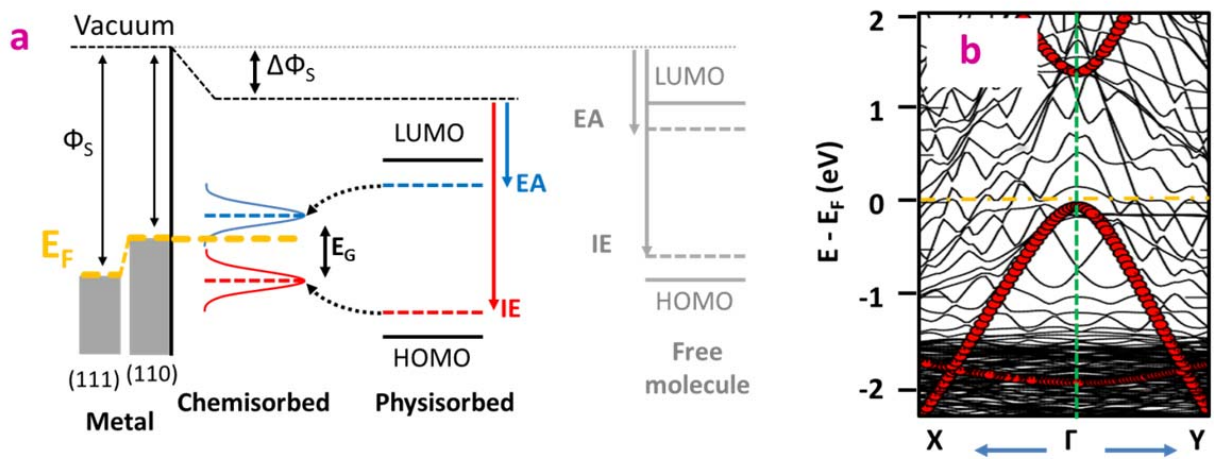

**Figure 5:** (a) Schematic representation of energy level alignments and band gap modifications in organic/metal interfaces. (b) DFT band-structure calculated enlarging the PPP-Cu(110) surface distance to 8 Å instead of 2.2 Å for the optimized structure. Red circles in the band-structure correspond to the projection on the first (molecular) layer.

**Methods**

**Sample preparation**

The experiments were carried out in ultrahigh vacuum at base pressures of $10^{-10}$ mbar or better. The Cu(110) crystal was prepared by repeated cycles of sputtering (Ar+,1 keV) and annealing (750 K). 1,4-dibromobenzene molecules (Sigma Aldrich, purity 98%) were sublimed onto the surface (held at room temperature), using a leak valve at partial pressures of $10^{-8}$ mbar. The PPP polymer presented here was obtained by reaching the saturation coverage of the surface. Under these conditions, a one monolayer thick organic layer was accumulated after t=9 min of deposition corresponding to 7 Langmuir. After deposition, the sample was post-annealed between 450 and 550 K for 5 min, to fully polymerize the system.[25]

**STM/STS and ARPES measurements**

Scanning tunneling microscopy and spectroscopy experiments were carried out using an Omicron LT-STM at a temperature of 5 K. STM images were recorded at constant tunneling current (0.2 nA) and constant bias voltage (applied to the sample). The dI/dV spectra were recorded in the open feedback loop mode ($V_{stab}$=2V) using the lock-in technique (peak to peak modulated voltage $V_{pp}$ = 30 mV, f=1100 Hz). Spectra were normalized by subtracting a background corresponding to the clean surface. ARPES experiments were performed at the CASSIOPEE beamline (synchrotron SOLEIL, Paris, France). The polymerization process and the orientation of the polymers were carefully analyzed using the characteristic LEED and XPS signatures (respectively shown in figure S1, S2, SI and ref. 25). The data presented in Figure 3 were measured at 30 K using a Scienta SES 200 high-resolution hemispherical electron analyzer, with linearly polarized 35 eV photons (for the experimental geometry see figure S5, SI)

**Tight-binding and first-principle band structure calculations**

The free-molecule orbitals and energy levels were obtained using a standard Hückel model. The hopping integral between two first neighbour carbon atoms was assumed to be constant equal to β. The Coulomb integral α was also assumed to be the same for all atoms and was adjusted to rigidly shift the band structure. STS differential conductivity has been obtained for a N=6 oligomer by taking the square–modulus of the TB wave functions as function of the quantized energies. Calculated ARPES intensities maps were obtained by taking the square modulus of the Fourier transform of the calculated orbitals, according to the method presented in refs. *39,41*. The calculated ARPES spectral weight shown in Figure 3.e was obtained by calculating orbitals of three chains of 20 phenyl rings with an inter-chain coupling constant β' of 0.15 eV.

DFT calculations were performed using the plane-wave pseudopotential code VASP[32-34]. We used GGA formulation to approximate the exchange-correlation functional potential proposed by Perdew, Burke, and Enzerhof (PBE). The exchange energy was described on the GGA level using the recently introduced optB86b functional. This approach yields a very good description of the vdW interactions between molecules and surfaces. The size of the unit cell was first obtained from the relaxation of a clean Cu(110) surface. The unit cell dimensions were 4.41519 x 12.7455 x 25.00 Å$^3$ ($\gamma$=54.736 degrees). The slab contained 40 copper atoms in an eight-layers. The five bottom layers were kept fixed during relaxation, at positions corresponding to the copper bulk. A vacuum region of at least 12 Å was maintained to avoid spurious interactions with periodic images. The electronic structure calculations were performed using a 6 x 2 x 1 Monkhorst-Pack grid to sample the Brillouin zone (this corresponds to 6 non-equivalent k points). The pseudopotentials were expressed within the projector augmented wave (PAW) scheme with an energy cut-off of 500 eV and a smearing $\sigma$=0.05 eV. Once the candidate structures were relaxed within 0.001 eV/Å, the corresponding (constant current) STM images were computed within the Tersoff-Hamann approximation[36], where the current at a given tip position above the sample is expressed as the integral of the density of states between the Fermi energy and the applied potential. The band structure plots were prepared using the converged Kohn-Sham orbitals during a non-self-consistent run along the four high-symmetry directions of the first Brillouin zone corresponding to the unit cell described above (each direction was sampled using sixty points). The band structures presented here also include plots of the bands projected on the top layer of the unit cell. The local density of states plots were projected on individual top layer atoms, as indicated in the figure legend. Additional DFT calculations have been carried out in the gas phase using Gaussian 09 at the B3LYP/631g(d) level (see section 8, SI).

**Acknowledgements**

This work is supported by the Conseil Franco-Québecois de Coopération Universitaire. D.F.P. and F.R. are supported by NSERC Discovery Grants as well as an FRQNT team grant and an MEIE project (collaboration with Belgium). F.R. acknowledges NSERC for an EWR Steacie Memorial Fellowship and Elsevier for a grant from Applied Surface Science. L.C. acknowledges partial salary support through a personal fellowship from FRSQ.

**Author contributions**

Y.F.R., D.M., F.R., J.L.D, G.C. and D.F.P. conceived the project. Y.F.R., G.V., L.C. and L.M. participated to the elaboration of the material. G.V., Y.F.R, M.S., M.D.G., G.G. did the sample preparation and ARPES measurements at the synchrotron. P.L.F. and F.B. provided help to do ARPES on the CASSIOPEE beamline. G.V. with the help of M.S. did the STS experiments under the guidance of B.K. Preliminary results have been obtained by YFR, LC, M.D.G. and G.C.. G.V. did the Tight-Binding modelisation of STS and ARPES data. L.L. and V.M. did the DFT calculations at surface. J.L.D. did the gas phase DFT calculations. YFR coordinated the work and wrote the manuscript with the help of G.V. All the authors commented critically on the manuscript.

**Corresponding Author**

*E-mail: yannick.fagot-revurat@univ-lorraine.fr